\documentclass[prl,superscriptaddress,preprint,endfloats,nopacs]{revtex4}

\usepackage{amsmath, amsthm, amssymb, pifont, wasysym}

\usepackage{graphicx}
\usepackage{bm} 
\usepackage{pifont}
\usepackage{dcolumn} 
\usepackage{color}
\definecolor{mygray}{gray}{0.5}
\newcommand {\etal}{\textit{et al}.}

\newcommand {\Bsatin}{$B_{\mathrm{sat,in}}$}
\newcommand {\Bsatout}{$B_{\mathrm{sat,out}}$}

\newcommand {\ECS}{EuCd$_2$Sb$_2$}

\newcommand {\MS}{Mn$_3$Sn}
\newcommand {\CSS}{Co$_3$Sn$_2$S$_2$}

\newcommand {\Ryx}{$R_{\mathrm{yx}}$}
\newcommand {\RyxAHE}{$R_{\mathrm{yx,AHE}}$}
\newcommand {\RyxAHEz}{$R_{\mathrm{yx,AHE}}^0$}

\newcommand {\sxyAHE}{${\sigma_{\mathrm{xy,AHE}}}$}

\begin{document}

\title{Observation of in-plane anomalous Hall effect \\associated with orbital magnetization}

\author{Ayano Nakamura}
\affiliation{Department of Physics, Tokyo Institute of Technology, Tokyo 152-8551, Japan}
\author{Shinichi Nishihaya}
\affiliation{Department of Physics, Tokyo Institute of Technology, Tokyo 152-8551, Japan}
\author{Hiroaki Ishizuka}
\affiliation{Department of Physics, Tokyo Institute of Technology, Tokyo 152-8551, Japan}
\author{Markus Kriener}
\affiliation{RIKEN Center for Emergent Matter Science (CEMS), Wako 351-0198, Japan}

\author{Yuto Watanabe}
\affiliation{Department of Physics, Tokyo Institute of Technology, Tokyo 152-8551, Japan}
\
\author{Masaki Uchida}
\email[Author to whom correspondence should be addressed: ]{m.uchida@phys.titech.ac.jp}
\affiliation{Department of Physics, Tokyo Institute of Technology, Tokyo 152-8551, Japan}


\begin{abstract}
For over a century, the Hall effect, a transverse effect under out-of-plane magnetic field or magnetization, has been a cornerstone for magnetotransport studies and applications. Modern theoretical formulation based on the Berry curvature has revealed the potential that even in-plane magnetic field can induce anomalous Hall effect, but its experimental demonstration has remained difficult due to its potentially small magnitude and strict symmetry requirements. Here we report observation of the in-plane anomalous Hall effect by measuring low-carrier density films of magnetic Weyl semimetal {\ECS}. Anomalous Hall resistance exhibits distinct three-fold rotational symmetry for changes in the in-plane field component, and this can be understood in terms of out-of-plane Weyl points splitting or orbital magnetization induced by in-plane field, as also confirmed by model calculation. Our findings demonstrate the importance of in-plane field to control the Hall effect, accelerating materials development and further exploration of various in-plane field induced phenomena.
\end{abstract}
\maketitle
\newpage

The Hall effect is the generation of a voltage transverse to an electric current under out-of-plane magnetic field or magnetization \cite{Hall1, Hall2, AHEreview}. This finds many applications such as sensors and opens up a variety of research areas along with theoretical formulation based on the Berry curvature \cite{AHEreview, AHEexp1, AHEexp2}, including the quantum Hall effect \cite{QHE} and further exotic Hall effects involving quasiparticles other than electrons \cite{phononHall, magnonHall}. In-plane Hall effect is defined as a transverse effect induced by in-plane magnetic field. So far many theoretical ideas for realizing this effect, or more precisely, in-plane anomalous Hall effect (AHE) have been proposed \cite{iAHEthe1_Q, iAHEthe2_Q, iAHEthe3_Q, iAHEthe4_Q, iAHEthe7_Q, iAHEthe8, iAHEthe5, iAHEthe6}, especially focusing on topological aspects of the band structure. Their main motivation is to elucidate whether it is possible to induce out-of-plane magnetization by in-plane field. However, experimental efforts to observe such a Hall signal are currently very limited \cite{iAHEexp1, iAHEexp2, iAHEexp3, iAHEexp4}.

One difficulty for observation of the in-plane AHE is its potentially small magnitude. As the magnitude is proportional to net magnetization, a ferromagnetically ordered state is desirable. But usually out-of-plane net magnetization does not couple to in-plane field, and for an in-plane ferromagnetic state, the out-of-plane component of spin magnetization should be zero. On the other hand, orbital magnetization induced by the orbital motion of electrons is also important when considering intrinsic AHE \cite{orbitalmagterm1}, as modernly formulated in connection with the Berry curvature \cite{orbitalmag_rev, orbitalmag1, orbitalmag2}. The orbital magnetization may not be aligned in the field direction even after the spin magnetization has saturated, although it is usually very small in the total magnetization.

The other difficulty for its observation is symmetry requirements. The Hall effect is always allowed by breaking of time-reversal symmetry under out-of-plane field $B_z$. Hall resistance $R_{yx}$, defined by the ratio of voltage $V_y$ to current $I_x$, simply appears reflecting a signed $z$ component of normalized field $\bm{B}/B$, as sketched in Fig. 1(a). In contrast, there are some additional symmetry requirements for the in-plane AHE \cite{iAHEsym}. Since this is a field-odd effect, for example, it is prohibited in a system with $C_{2}$, $C_{4}$, or $C_{6}$ rotational symmetry along the $z$ direction. $R_{yx}$ thus appears with one-fold or three-fold rotational symmetry including its sign, as exemplified in Fig. 1(b). It is necessary to carefully examine other effects as induced by sample misalignment, especially in the case of one-fold component. We also note that the in-plane AHE is essentially different from so called planar Hall effect, a field-even effect originating in anisotropy of magnetoresistance \cite{PHE, iAHEsym}.

Here we report experimental observation of the in-plane AHE in magnetic Weyl semimetal films. This appears robust to changes in the out-of-plane field component and exhibits clear three-fold symmetry for changes in the in-plane one. In simple magnetic Weyl semimetals \cite{Weyl1, Weyl2}, AHE is enhanced upon approaching the Weyl point energy, and where it is proportional to splitting of the paired Weyl points \cite{Weyl_AHE}. At the same time, the orbital magnetization, which appears along the out-of-plane field direction as illustrated in Fig. 1(c), is also proportional to the splitting of the Weyl points \cite{orbitalmagterm2}. Such out-of-plane Weyl points splitting or orbital magnetization can be induced also by the in-plane field in principle, as sketched in a simplified form in Fig. 1(d). 

Aiming to observe the in-plane AHE, we focus on low-carrier density films of magnetic Weyl semimetal {\ECS}. In {\ECS}, Cd $5s$ and Sb $5p$ bands are inverted to form a simple but topological band structure, which hosts a few pairs of Weyl points near the Fermi energy in forced ferromagnetic states \cite{ECS_film, ECS_Weyl, ECS_nMC}. {\ECS} has a trigonal crystal structure with space group $P\bar{3}m1$ \cite{ECS_crystal}, where magnetic Eu layers and itinerant Cd$_{2}$Sb$_{2}$ blocks are alternately stacked along the out-of-plane $c$-axis, as shown in Fig. 2(a). There are $C_3$ rotational symmetry along the $c$-axis, $C_2$ rotational symmetry along the in-plane $a$- and its equivalent axes, and mirror symmetry perpendicular to them. At zero field, {\ECS} shows so-called A-type antiferromagnetic ordering, in which the spin magnetic moments are ferromagnetically ordered on a triangular lattice $ab$-plane and antiferromagnetically stacked along the $c$-axis \cite{ECS_magnetic}.

[001]-oriented {\ECS} single-crystalline thin films were grown by molecular beam epitaxy on ($111$)A CdTe substrates in an EpiQuest RC1100 chamber \cite{ECS_film, CdTe_etching, Supplemental}. Importantly, we have succeeded in fabricating single-domain films with only negligibly small amount of $60^{\circ}$ rotated domains \cite{Supplemental}, which is essential for detecting in-plane AHE in this system. Hall resistance {\Ryx} was measured with a standard four probe method on a Hall bar. Low-temperature measurements up to 9 T were performed using a Cryomagnetics cryostat system equipped with a superconducting magnet. Their field angle dependences were measured using a sample rotator in a Quantum Design Physical Property Measurement System (PPMS). The transverse Hall voltage was measured by feeding an electric current primarily along the {\ECS} [$110$] direction, with changing direction of the applied magnetic field, from out-of-plane [$001$] to in-plane [$110$] with an angle $\theta$ (measured from the [$001$] direction), and also within the ($001$) plane with an angle $\varphi$ (measured from the [$110$] direction). Magnetization curves up to 7 T were measured using a superconducting quantum interference device magnetometer in a Quantum Design Magnetic Property Measurement System (MPMS). Here we constructed an effective model considering the four valence bands consisting of Sb $^2$P$_{3/2}$ orbitals \cite{Supplemental}. To calculate the band structure under the magnetic field, the magnetic field dependences of $m_{x,y}$ are assumed to be $m_{x,y}=h_{x,y}$ for $h=\sqrt{h_x^2+h_y^2 }\leq 1$ and $m_{x,y}=h_{x,y} /h$ for $h>1$. The Hall conductivity was calculated using Kubo formula. To calculate the Hall conductivity for the hole band, we assumed that the Hall conductivity at the charge neutral point is zero.

Figure 2(b) shows relation between the anomalous Hall resistance {\RyxAHEz} and the magnetization $M$ measured for an out-of-plane magnetic field ($\theta = 0 ^{\circ}$). As previously reported for {\ECS} bulks \cite{ECS_magnetic, ECS_Weyl} and also films \cite{ECS_film}, the spin magnetic moments are gradually canted from the A-type antiferromagnetic ordering with increasing the field, and forcedly aligned along the $c$-axis above the out-of-plane saturation field {\Bsatout} = 3.5 T. Here {\RyxAHEz} is extracted by subtracting the field-linear term in a conventional way so that {\RyxAHEz} becomes flat above the saturation field. {\RyxAHEz} exhibits a non-monotonic field dependence with a peak structure below {\Bsatout}. Such AHE not proportional to the magnetization has been observed in low-carrier density Weyl semimetals or their candidates hosting Berry curvature hot spots \cite{ETO_AHE, ECS_film, ECA_THE, Eu_future_ECA}. It has been theoretically shown that formation and energy shift of the Weyl points across the Fermi energy accompanied with the Zeeman-type band splitting results in non-monotonic modulation of the Berry curvature and intrinsic AHE \cite{ETO_AHE}. Since no other magnetic orderings such as non-coplanar spin arrangement have been confirmed in {\ECS} \cite{ECS_magnetic, ECS_Weyl}, the peak structure in {\RyxAHEz} is also ascribed to the non-monotonic change in the Berry curvature during the magnetization process. 

Next we present {\RyxAHEz} and $M$ taken for an in-plane magnetic field ($\theta = 90 ^{\circ}$) in Fig. 2(c). In the case of in-plane field, a spin flop transition may occur at about $0.2$ T, depending on the direction of the spin magnetic moments relative to the in-plane field in three types of antiferromagnetic domains, and then the spin magnetic moments are fully aligned along the in-plane field direction above the in-plane saturation field {\Bsatin} = 1.8 T \cite{ECS_magnetic}. Note that {\RyxAHEz} exhibits a finite signal with a clear peak structure even for the in-plane field. This {\RyxAHEz} is not induced by the out-of-plane component of possibly misaligned in-plane field, as evidenced by the following observations. Figure 2(d) shows {\RyxAHEz} curves when the magnetic field is rotated from $\theta = 0 ^{\circ}$ to $180 ^{\circ}$ across the in-plane direction ($\theta = 90 ^{\circ}$). At first, the peak structure, which appears not proportional to the out-of-plane magnetization below the saturation field, becomes smaller upon increasing $\theta$ and disappears at $\theta = 75^{\circ}$. With further increasing $\theta$, a peak with opposite sign emerges at $\theta = 90 ^{\circ}$. Due to this component, data taken at a pair of angles $\theta = 90^{\circ}\pm\theta'$ are not antisymmetric as confirmed in the data at $\theta = 75^{\circ}$ and $105^{\circ}$ ($\theta' = 15^{\circ}$). Figure 2(e) shows {\Ryx} data taken for finer $\theta$ steps around $\theta = 90 ^{\circ}$. The peak structure robustly appears without depending on the out-of-field component, even when the magnetic field crosses \textcolor{black}{the in-plane direction and the signs of the ordinary Hall effect (OHE) and the out-of-plane AHE are inverted. It is also worth mentioning that the signs of slope and intercept in {\Ryx} appear to be reversed at slightly different angles; the slope of {\Ryx} becomes zero between $\theta = 89 ^{\circ}$ and $90 ^{\circ}$, while the intercept in linear extrapolation of {\Ryx} from above {\Bsatin} to zero field becomes zero around $\theta = 92 ^{\circ}$. As discussed in detail in the supplemental material \cite{Supplemental}, these observations also indicate that there is a unique AHE component with finite slope and intercept, which is not induced by the out-of-plane magnetic field but the in-plane one, namely, the in-plane AHE component.}

Figure 3(a) shows in-plane field dependence of AHE taken at different azimuthal angles $\varphi = 0^{\circ}$, $30^{\circ}$, and $90^{\circ}$ for the same {\ECS} film at $2$ K (sample A). Here {\RyxAHE} for the in-plane field is extracted by subtracting the one-fold component which consists of the out-of-plane OHE and AHE, with referring to $\varphi$ scans \cite{Supplemental}, and this indeed reflects intrinsic contribution of the in-plane AHE. Since {\RyxAHE} may be variable even above {\Bsatin} reflecting change in the orbital magnetization, it is necessary to carefully evaluate it based on the $\varphi$ scans, different from the conventional subtraction procedure for {\RyxAHEz} in Fig. 2. When the in-plane field is applied parallel to the current at $\varphi = 0 ^{\circ}$, no evident {\RyxAHE} is observed above {\Bsatin}. This is consistent with the symmetry conditions of absence of the in-plane AHE, namely, $C_2$ axis along the field direction and/or mirror plane perpendicular to it. At $\varphi = 30 ^{\circ}$ and $90^{\circ}$, on the other hand, {\RyxAHE} exhibits clear signals with opposite signs, characterized by the peak structure below {\Bsatin} and unsaturated linear increase even after saturation of the spin magnetization above {\Bsatin}. As confirmed by model calculation later, this unsaturated behavior can be explained by continued separation of the Weyl points in the $k_z$ direction associated with the Zeeman-type band splitting.

To check reproducibility, we have also examined the azimuthal angle dependence on a different {\ECS} film (sample B), with rotating the current direction $30$ degrees from the case of sample A. Figure 3(b) compares {\RyxAHE} measured at $\varphi' = 0^{\circ}$ for sample B to $\varphi = 0^{\circ}$ for sample A, where $\varphi'$ is measured from the [$120$] current direction. Clear in-plane AHE signals appear at $\varphi' = 0^{\circ}$ as in $\varphi = 30^{\circ}$, not $\varphi = 0^{\circ}$, in sample A. This confirms that the in-plane AHE emerges depending on the field direction relative to the crystal orientation, not the current direction. \textcolor{black}{The difference in the peak amplitudes between samples A and B is probably attributed to the difference in the hole density or the Fermi level.}

Figure 3(c) shows $\varphi$ scans of {\RyxAHE} measured by changing $\varphi$ in the plane at various constant fields for sample A. Reflecting the crystal symmetry, {\RyxAHE} in the $\varphi$ scans exhibits clear three-fold symmetry for all the magnetic fields even before subtraction of the one-fold out-of-plane component \cite{Supplemental}. For example at $\varphi = 30^{\circ}$, {\RyxAHE} exhibits a positive maximum and turns negative just below {\Bsatin} = 1.8 T, and then continues to decrease above {\Bsatin}, as also confirmed in the field scan of {\RyxAHE} in Fig. 3(a). Above {\Bsatin}, {\RyxAHE} exhibits a node every $60$ degree at $\varphi$ = $0^{\circ}$, $60^{\circ}$, $\cdots$, $300^{\circ}$. Below {\Bsatin}, on the other hand, these node positions are slightly shifted by about 5 degrees at maximum. This is because strictly speaking, canted A-type antiferromagnetic ordering for example at $\varphi = 0^{\circ}$ breaks the $C_2$ symmetry along the [110] direction, resulting in a finite in-plane AHE signal below {\Bsatin} as seen for $\varphi$ = $0^{\circ}$ in Fig. 3(a). Its detailed behavior at low fields would be interesting for future studies. Figure 3(d) illustrates the $\varphi$-dependent {\RyxAHE} above {\Bsatin} on the {\ECS} ($001$) plane. Again, the in-plane AHE appears fully following the crystal symmetry of {\ECS}. We also emphasize that the observed three-fold symmetry in the in-plane AHE is in stark contrast to the field-even two-, four-, or six- fold symmetry in the conventional planar Hall effect.

Figure 4(a) shows in-plane field dependence of anomalous conductivity {\sxyAHE} for $\varphi = 0^{\circ}$ and $90^{\circ}$, calculated based on the effective model of {\ECS} near the Fermi energy. While {\sxyAHE} is zero at $\varphi = 0^{\circ}$, it takes finite values below and above the saturation field {\Bsatin} ($h=1$) at $\varphi = 90^{\circ}$. Non-monotonic behavior as also experimentally observed in the magnetization process is similar to one calculated for out-of-plane AHE \cite{ETO_AHE}. This indicates that formation and energy shift of the Weyl points accompanied with the Zeeman-type band splitting is also induced by the in-plane field, causing non-monotonic change in the Berry curvature. In addition, the unsaturated linear increase above the saturation field is also theoretically confirmed. This indicates that separation of the paired Weyl points in the $k_z$ direction continues with the Zeeman-type band splitting, which can also be captured in a more intuitive picture that the out-of-plane orbital magnetization keeps increasing with increasing the in-plane field. As confirmed in Fig. 4(b), the out-of-plane Weyl points splitting or orbital magnetization is induced certainly reflecting the crystal symmetry. It shows in-plane three-fold symmetry with a node at $\varphi$ = $0^{\circ}$, $60^{\circ}$, $\cdots$, $300^{\circ}$, also consistent with the observation.

Figures 4(c)-4(e) schematically summarize the electronic structure change in {\ECS}, where the Weyl points split in different ways depending on the magnetic field direction. In the antiferromagnetic ground state, paired Weyl points can be seen almost degenerate, forming Dirac points at $(0, 0, \pm k_\mathrm{D})$, since a gap opened by the antiferromagnetic ordering is negligibly small \cite{ECS_Weyl}. When a magnetic field is applied along the in-plane [$110$] direction ($\varphi = 0^{\circ}$), the spin magnetic moments are forcedly aligned along the field direction above {\Bsatin} \cite{ECS_magnetic}. In this forced ferromagnetic state, the Weyl points split symmetrically by $\pm\Delta \bm{k}$, roughly along the direction of magnetic field, as shown in Fig. 4(d). Similar splitting occurs when the field is applied along the [$\bar{1}10$] direction ($\varphi = 90^{\circ}$). However, in the [$\bar{1}10$] case, the Weyl points split asymmetrically due to trigonal warping. As demonstrated by the above model calculation, this out-of-plane splitting causes finite out-of-plane Berry curvature and anomalous Hall conductivity. This in-plane AHE can be simply interpreted as being associated with the change in the out-of-plane orbital magnetization, because the spin magnetization is always aligned along the in-plane field direction.

We have succeeded in observing distinct in-plane AHE in {\ECS} films, which can be understood in terms of out-of-plane orbital magnetization induced by in-plane field. The trigonal $\bar{3}m$ structure without $60^{\circ}$ rotated domains is suitable for demonstrating the in-plane AHE. The symmetry arguments are similarly applicable when the spin ordering is considered for the forced ferromagnetic state at each in-plane field angle. Experimentally, the three-fold in-plane AHE component can be easily distinguished from the one-fold out-of-plane OHE and AHE components. In addition to satisfying symmetry requirements, {\ECS} is a simple Weyl semimetal system with low carrier density, where the Weyl points are located near the Fermi energy and form large Berry curvatures over the relatively small Fermi surface \cite{ECS_Weyl, ECS_film}. Therefore the anomalous Hall conductivity originating in the orbital magnetization is enhanced \cite{orbitalmagterm1, orbitalmagterm2, Weyl_AHE}, and the in-plane AHE is observed with change in the out-of-plane orbital magnetization by the in-plane field.

While observation of the in-plane AHE is expected in various other systems such as those hosting Weyl points, it has not yet been reported in case of the well-known magnetic Weyl semimetals {\MS} \cite{Mn3Sn} and {\CSS} \cite{Co3Sn2S2}. To observe the in-plane AHE, it is first necessary to choose a proper crystal plane which satisfies the symmetry requirements. For example, the plane must not have out-of-plane $C_2$ axis nor in-plane mirror plane \cite{iAHEsym}. Symmetry of the wave functions is important also in terms of magnitude of the in-plane AHE. This is because the order $n$ in cross terms like $k_y^n\sigma_z$, which reflects warping of the band structure based on the symmetry, affects the coupling strength between the in-plane field and the out-of-plane magnetization. It may be also favorable when the Weyl points are located close to the Fermi level, but as in the case of the out-of-plane AHE, they are not necessarily required if a system hosts Berry curvature hot spots.

Lastly, our findings highlight the importance of the in-plane field to control the Hall signal, triggering the development of materials which exhibit in-plane AHE, possibly with greater intensity or at higher temperatures. As theoretically examined \cite{iAHEthe1_Q, iAHEthe2_Q, iAHEthe3_Q, iAHEthe4_Q, iAHEthe7_Q}, demonstration of its quantization may be another important direction. As also seen in the recent discoveries of non-linear Hall effect \cite{nonlinearHall} and crystal Hall effect \cite{crystalHall}, even more rich physics will be included in the in-plane Hall effect, by relating the Berry curvature to many degrees of freedom in solids. It may be also worth exploring variation of in-plane Hall effects involving quasiparticles other than electrons \cite{phononHall, magnonHall}.

\begin{acknowledgments}
This work was supported by JST FOREST Program Grant Number JPMJFR202N, by JSPS KAKENHI Grant Numbers JP22K18967, JP22H04471, JP21H01804, JP22H04501, JP22K20353, JP23K13666 from MEXT, Japan, and by 2022 Yoshinori Ohsumi Fund for Fundamental Research, Japan.
\end{acknowledgments}

\newpage

\renewcommand{\baselinestretch}{1.3}\normalsize
\begin{figure}
\begin{center}
\includegraphics[width=16.5cm]{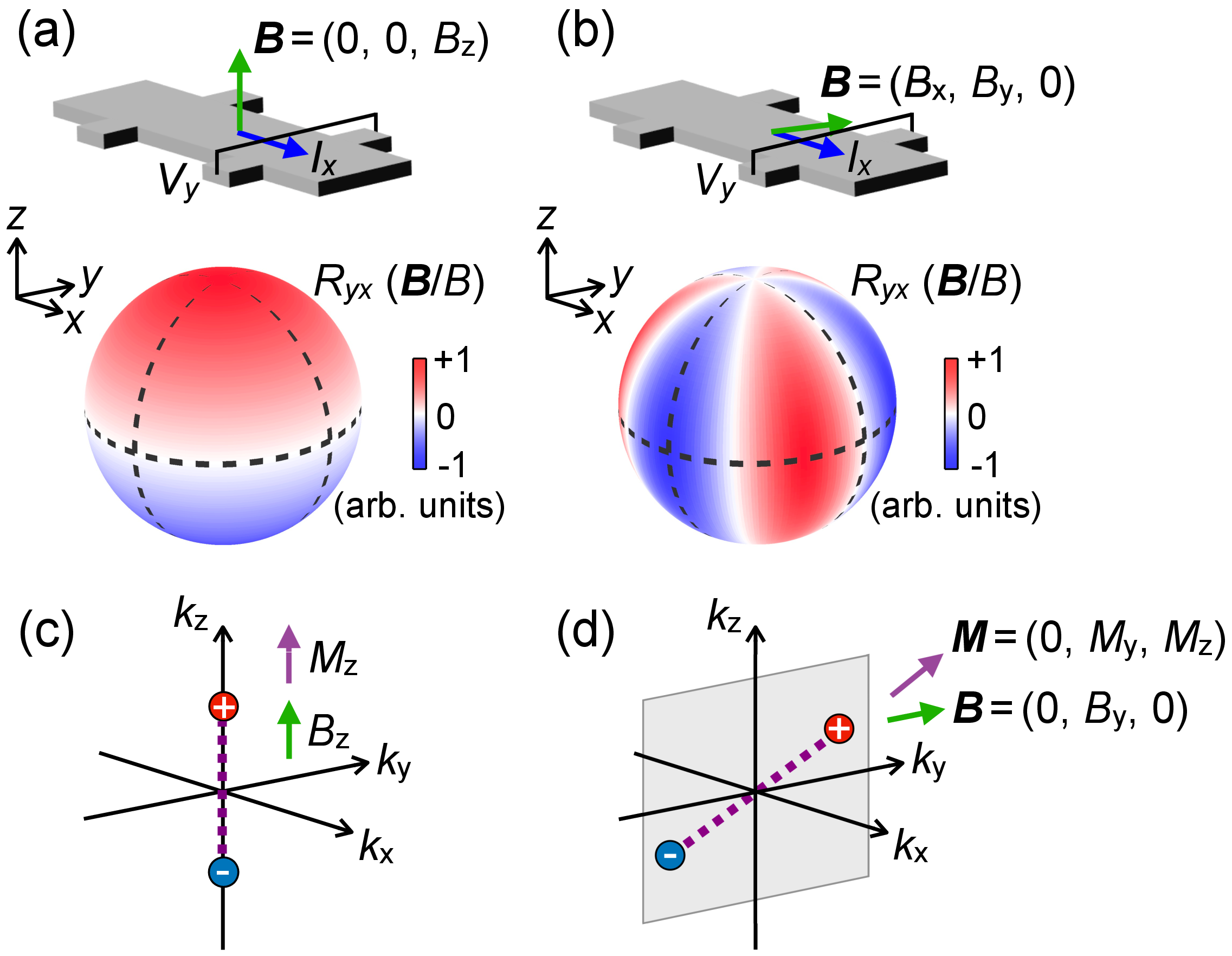}
\caption{
	Out-of-plane and in-plane Hall effects.
	Schematic illustrations of the measurement configuration and representative field direction dependence for (a) out-of-plane and (b) in-plane field induced Hall effects. In the case of in-plane Hall effect, Hall resistance {\Ryx} is induced by in-plane field components $B_\mathrm{x}$ and/or $B_\mathrm{y}$ with following one-fold or three-fold rotational symmetry along the $z$ direction. Simplified understanding of (c) out-of-plane and (d) in-plane anomalous Hall effects in term of Weyl points splitting or orbital magnetization in Weyl semimetals.
}
\label{fig1}
\end{center}
\end{figure}

\begin{figure}
\begin{center}
\includegraphics[width=16.5cm]{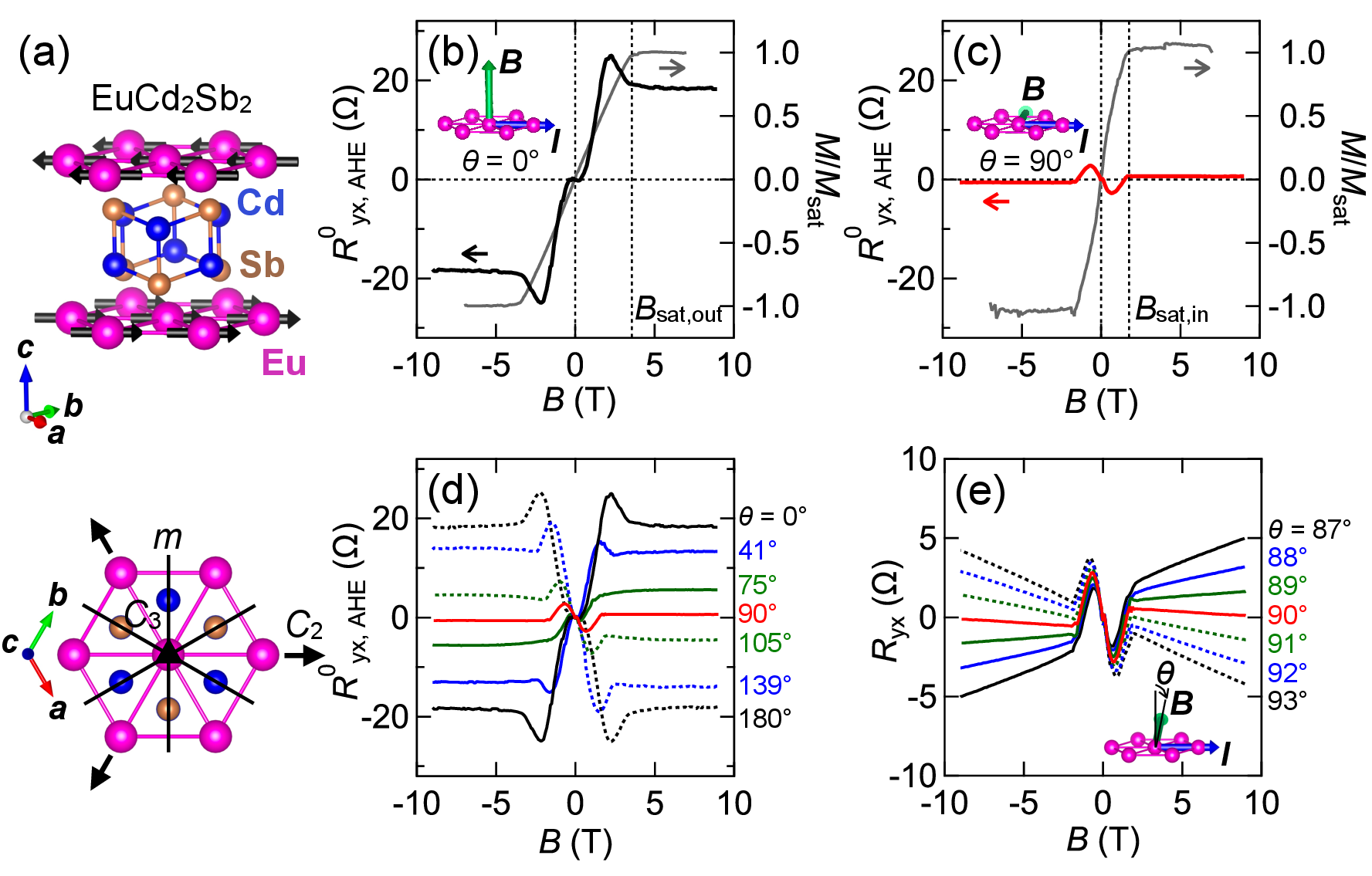}
\caption{
	Anomalous Hall effect induced by in-plane field.
	(a) {\ECS} crystal structure with antiferromagnetic spin ordering at the ground state. Fundamental symmetry elements are also shown in the bottom. Anomalous Hall resistance {\RyxAHEz} and magnetization $M$ of a {\ECS} film with hole density of $2.9 \times 10^{19}$ cm$^{-3}$ (sample A), measured for magnetic fields applied (b) out-of-plane and (c) in-plane at 2 K. (d) {\RyxAHEz} taken for various field direction angles $\theta$, from out-of-plane ($\theta = 0 ^{\circ}$) through in-plane ($\theta = 90 ^{\circ}$) to the opposite out-of-plane ($\theta = 180 ^{\circ}$). Data taken at pairs of angles $\theta = 90 ^{\circ}\pm\theta'$ across the plane are indicated by the same color. (e) Detailed $\theta$ dependence of {\Ryx} around $\theta = 90 ^{\circ}$.
}
\label{fig2}
\end{center}

\end{figure}

\begin{figure}
\begin{center}
\includegraphics[width=16.5cm]{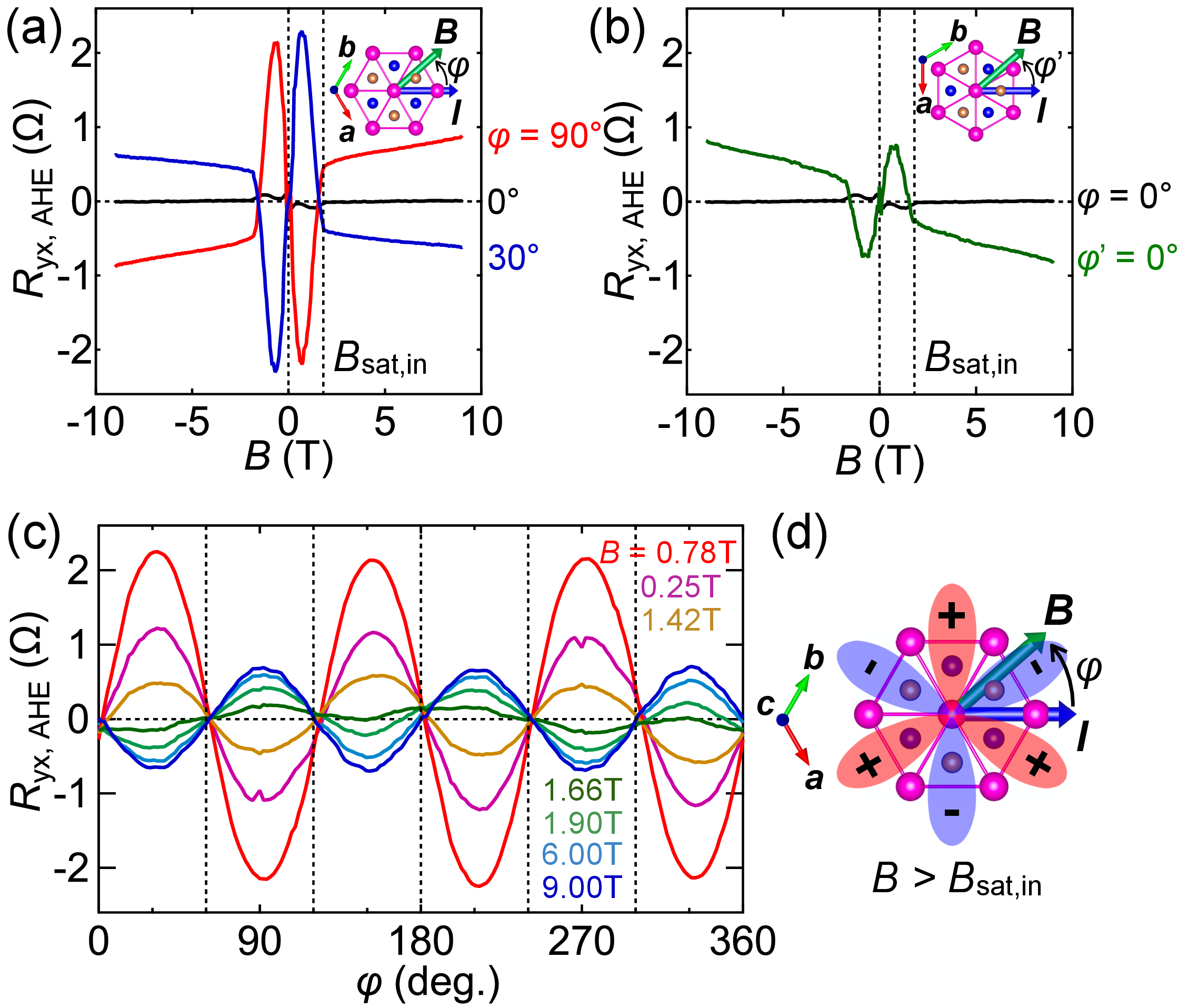}
\caption{
	Three-fold rotational symmetry of in-plane AHE.
	(a) Anomalous Hall resistance {\RyxAHE} of the {\ECS} film (sample A), measured for in-plane fields with different azimuthal angles $\varphi = 0^{\circ}$, $30^{\circ}$, and $90^{\circ}$ at 2 K. (b) {\RyxAHE} of another {\ECS} film with hole density of \textcolor{black}{$3.8 \times 10^{19}$ cm$^{-3}$} (sample B) measured at $\varphi' = 0^{\circ}$, compared with the result at $\varphi = 0^{\circ}$ in sample A. (c) {\RyxAHE} taken with continuously changing $\varphi$ at various magnetic fields for sample A at 2 K, exhibiting clear three-fold symmetry reflecting the crystal structure. (d) Schematic illustration of the $\varphi$-dependent in-plane AHE above {\Bsatin}.
} 
\label{fig3}
\end{center}
\end{figure}

\begin{figure}
	\begin{center}
\includegraphics[width=15.5cm]{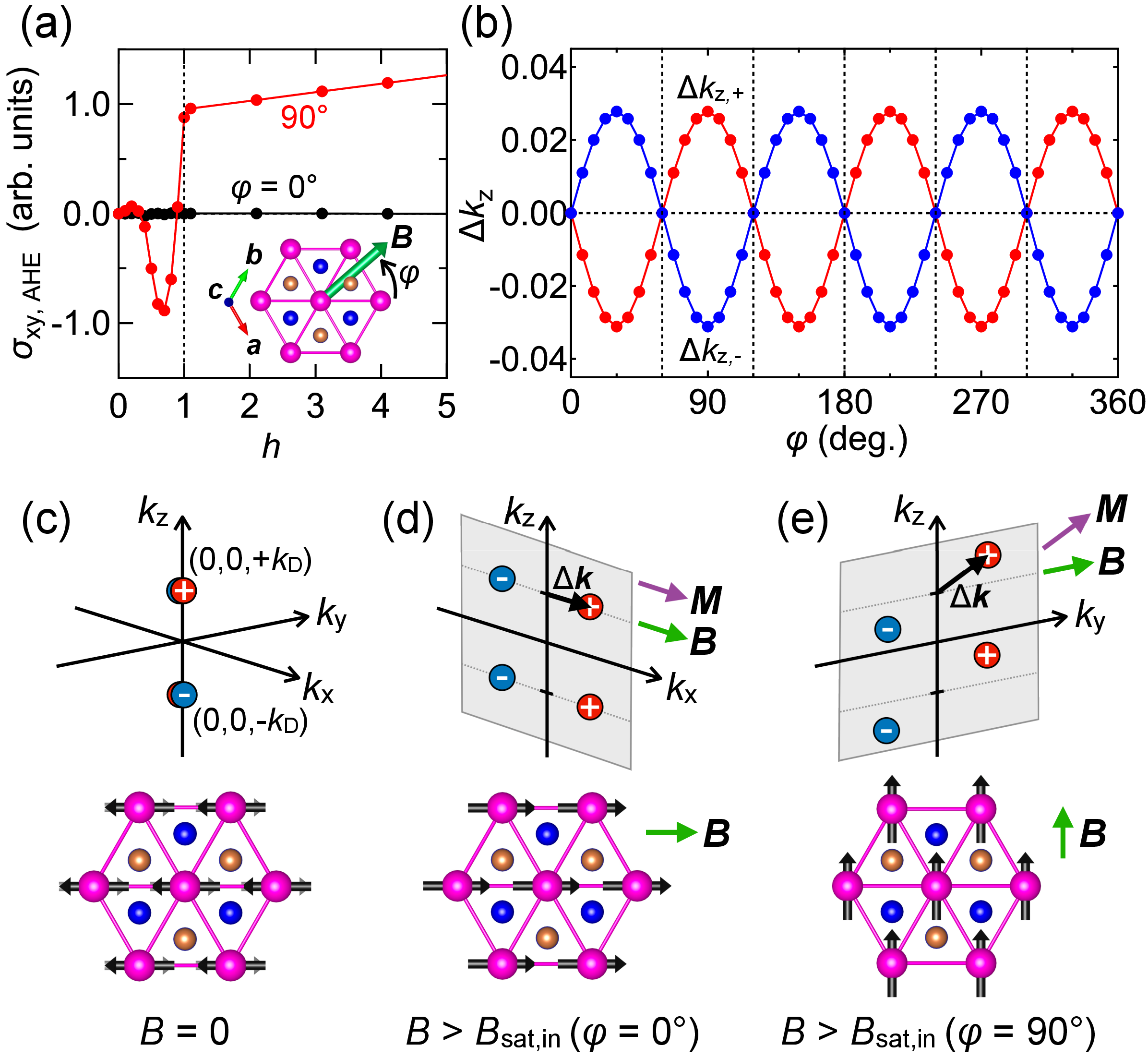}
\caption{
	In-plane AHE calculated for the magnetic Weyl semimetal.
	(a) In-plane field dependence of the anomalous Hall conductivity {\sxyAHE}, calculated for a model of magnetic Weyl semimetal {\ECS} at $\varphi = 0^{\circ}$ and $90^{\circ}$. Above the saturation field {\Bsatin} ($h=1$), the spin magnetic moments are fully aligned along the field direction. (b) $\varphi$ dependence of the $k_{\mathrm{z}}$ shift, calculated for paired Weyl points ($k_{\mathrm{z}}>0$) with positive or negative chirality at $h=1$. Schematic Weyl points splitting in (c) the A-type antiferromagnetic ground state and in the forced ferromagnetic states above {\Bsatin} for (d) $\varphi = 0^{\circ}$ and (e) $90^{\circ}$.
} 
\label{fig4}
	\end{center}
\end{figure}

\clearpage

\end{document}